# Freeform high-speed large-amplitude deformable Piezo Mirrors

Matthias C Wapler, Jens Brunne and Ulrike Wallrabe
University of Freiburg, Germany
E-mail: wallrabe@imtek.uni-freiburg.de

**ABSTRACT**

We present a new type of tunable mirror with sharply-featured freeform displacement profiles, large displacements of several 100µm and high operating frequencies close to the kHz range at 15mm diameter. The actuation principle is based on a recently explored "topological" displacement mode of piezo sheets. The prototypes presented here include a rotationally symmetric axicon, a hyperbolic sech-icon and a non-symmetric pyram-icon and are scalable to smaller dimensions. The fabrication process is economic and cleanroom-free, and the optical quality is sufficient to demonstrate the diffraction patterns of the optical elements.

## INTRODUCTION

### Background
Adaptive mirrors are used in many applications, for example for wavefront correction in telescopes [1]. For specific applications, however, varifocal elements with a pre-determined displacement profile will deliver better performance and lower cost. While varyfocal elements with spherical or parabolic profiles have been in use for a long time, producing elements with other profiles, such as axicons has been difficult, and problematic at small curvature radii [2].

### Design Concept
Our new approach is to deposit a reflective layer on top of a new kind of in-plane polarized piezo actuator in which the deflections arise solely from inhomogeneous in-plane deformations of a piezo film that cause the film to "pop out" of the plane. The out-of-plane displacement $h(r)$ of such actuators with a radial electric field E(r) and piezo coefficients $d_{31}$ and $d_{33}$ was found to be [3]

$$(\partial_r h(r))^2 = 2(d_{33}-d_{31})|E(r)| - 2d_{31}r\,\partial_r|E(r)|, \quad (1)$$

solely based on geometric considerations, ignoring forces and bending moments. The electric field in the piezo film is created by alternating interdigitated electrodes on one side of the piezo film, with an appropriate spacing to create the field distribution.

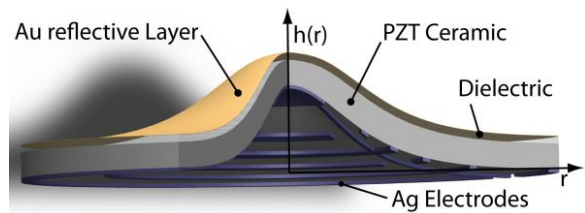

Fig. 1: Schematic of the displaced tunable mirror

In order to not disturb the electric fields in the actuator and to create a plane surface, the reflective layer is supported on a dielectric layer on top of the piezo as shown in fig. 1.

## MATERIALS AND FABRICATION

As a basis, we use 120µm thin PZT ceramic discs (€0.4) with 25mm diameter and double-sided silver electrodes from acoustic transducers. In the first step, the electrode is removed completely on one side, and structured with the electrode spacing obtained from eq. (1) on the other side, using a rapid-prototyping UV laser (a). The PZT is subsequently cleaned in weak ultrasound and depolarized at 400°C. At this stage (b), the films were uneven to several 100µm.

To produce the reflective layer and level the surface, we first deposit 120µm gold without adhesion layer on an auxiliary glass substrate. For the dielectric, liquid polyurethane (PU) resin is dispensed on top of the gold layer, and the PZT films are dropped inside a vacuum chamber (c) with their free side onto the PU resin. After curing the PU to D80 hardness, the mirrors are removed from the glass substrate using a vacuum chuck (d) and the perimeter is cut using a UV laser (e). Finally, they are weld-glued on top of a soft PU ring (or 4 pillars for the pyramid-mirrors) and polarized at a maximum field strength of 700V/mm. The final thickness of the dielectric layer was varying from 20µm to 80µm and is not accounted for in (1), as are also not the supporting ring, the connecting radial electrodes and the contact pads.

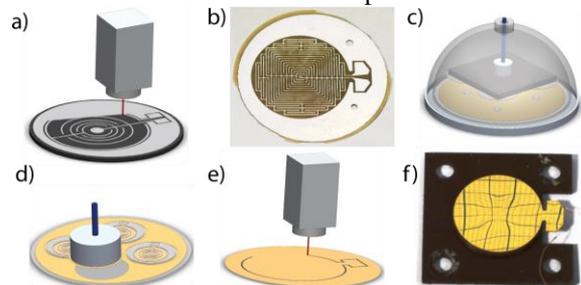

Fig. 2: Fabrication process and photos of a pyram-icon PZT sheet and a final displaced sech-icon.

## MECHNICAL CHARACTERIZATION

### Static Behavior
The mechanical characterization was performed by scanning the surface with a laser triangulation sensor. A first scan of the surface before polarization (fig. 3a) showed a surface unevenness in the range of ±10µm over 14mm, and local curvatures corresponding to a focal length down to 250mm. A similar pattern appears in the deviation from rotational symmetry (3b) at an actuation of 700V/mm, so the actuation itself does not add significantly to the unevenness.

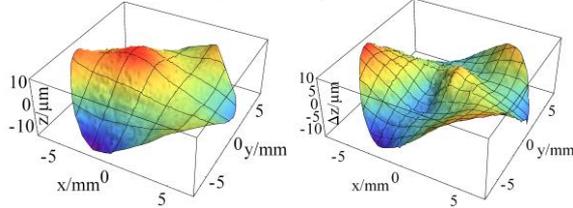

Fig. 3: Surface unevenness initially and at deflection. The apparent roughness is the measurement noise.

Fig. 4 compares the profiles at maximum deflection to the scaled target for an axicon and a sech-icon. There is a very good agreement up to some rounding due to bending moments and a shift in the length scale of the secans hyperbolicus from 1.5 to 2.0mm.

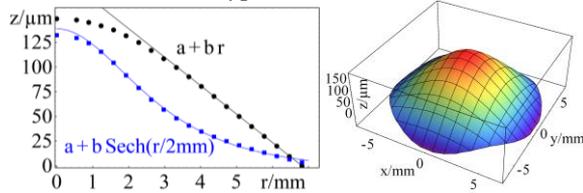

Fig. 4: Radial profiles of the axicon and sech-icon compared to the target and surface of the pyram-icon.

**Dynamic Behavior**

The first resonance occurs always between 0.5 and 1kHz, with the usual ratio of the higher resonance frequencies of a circular disk. The different deflection profiles couple at different strength to the resonances. Both the frequencies and the hysteresis vary highly for different mirrors, probably due to the varying material thickness and the viscoelasticity of PU.

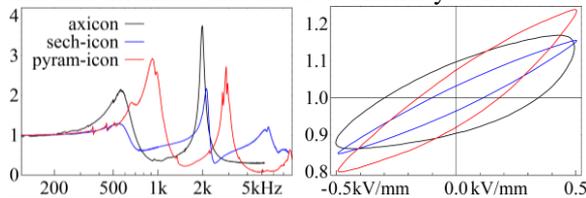

Fig. 5: Resonance spectrum, displacement around the remanent pre-deflection, both normalized.

**OPTICAL CHARACTERIZATION**

The optical characterization was performed at an upwards-deflected sech-icon which we simulated with a discrete Hankel transformation [4] and a downward pyram-icon that we simulated with an FFT.

The sech-icon produces a hollow beam with a sharp outer edge and an outer ring pattern. Illuminated with an aperture, of 7mm, the intensity profile reproduces the simulation very well, up to some structures at the center that are an artefact of the electrodes on the rear side of the PZT substrate and a rotational asymmetry. Qualitatively, also the pyram-icon reproduces the effects of a rounded pyramid-mirror, with an outer quadratic ring pattern and an inner interference grid.

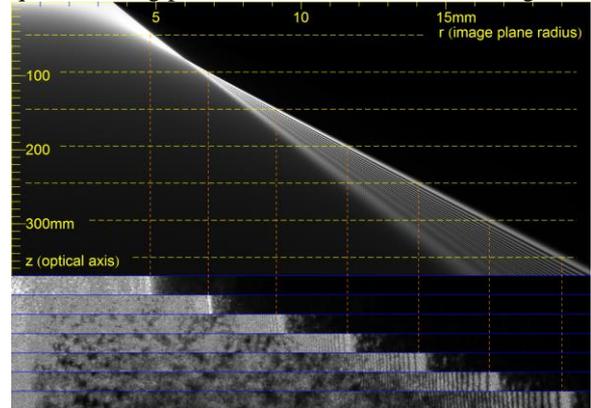

Fig. 6: Simulated intensity distribution of a sech-icon at 100µm deflection; measurements at -0.25kV/mm.

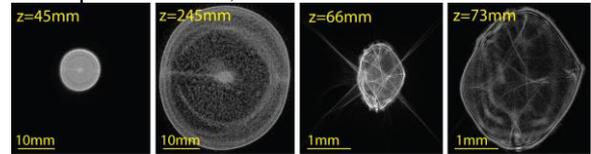

Fig. 7: Sech-icon and pyram-icon intensity images.

**CONCLUSIONS**

We have validated a new principle to produce tunable optical reflective elements that can operate at high frequencies and large deflections with a high design freedom of the deflection profile. In particular the possible angles of the surfaces do not decrease as we scale down the elements, so the focal length scales scale down with the element size, as will the resonance frequency inversely scale up into the kHz range. As we used very rudimentary PZT sheets and support structures, the surface quality was not in the range required for actual applications, but still we showed the first tunable sech-icon and pyram-icon, demonstrating the potential of this type of device.

**Acknowledgments**
This research was financed by the Baden-Württemberg Stiftung gGmbH.